\documentclass[prl,showpacs,twocolumn,aps]{revtex4}
\usepackage{graphicx}
\usepackage{epsfig}

\begin{document}

\title{ Universality  of Entropy Scaling in 1D Gap-less Models }
\author{ V.E.\ Korepin }
\affiliation{C.N.\ Yang Institute for Theoretical Physics,\\
 State University of New York at Stony Brook, Stony Brook, NY 11794-3840
  {\footnote {korepin@insti.physics.sunysb.edu}} }
\date{\today}

\begin{abstract}
We consider critical models in one dimension.
We study the ground state in thermodynamic limit [infinite lattice].
Following  Bennett,  Bernstein, Popescu, and Schumacher, we use the 
entropy of a sub-system as  a measure of  entanglement.
We calculate the  entropy of a part of the ground state. At zero temperature
 it describes entanglement of this part with
 the rest of the ground state.
We obtain an explicit  formula for the entropy of the subsystem
at low temperature.
At zero temperature  we reproduce a logarithmic
 formula of Holzhey,  Larsen and  Wilczek.
Our derivation is based on   the second law of thermodynamics. 
The entropy of a subsystem is calculated explicitly  for   Bose gas with delta interaction, the  Hubbard
 model and spin chains with arbitrary value of spin.

\end{abstract}


\maketitle

Entanglement is an important resource for quantum computation 
\cite{ben,lind,nlind,lmez,zimb, ekert,cirac,lloyd, pach,joz,rich,wint,christ,ved,bose, sougato}. 
Recently entanglement was studied in detail in  ground state of different 
physical models \cite{plenio, zan, eis, keat,mez,  cardy, pop, sal,lat}.

Conformal field theory \cite{bpz} is  useful for the description of low 
temperature behavior of gap-less models in one space and one time dimensions.
Critical models are classified by  a central charge $c$ of corresponding 
Virasoro algebra. A definition of the  central charge is given the the 
 Appendix. 
Conformal field theory is  closely related to Luttinger liquid \cite{h} approach.
We are interested in  specific entropy $\tt s$ [entropy per unit length].
Let us start with  specific heat $C=T d{\tt s}/dT$.
Low temperature behavior  was obtained in \cite{bcn,a}:  
$$C={\pi T c k^2_B \over 3 \hbar v}.$$
 Here  $c$ is a central charge  and $v$
is Fermi velocity. We put both Plank  and Boltzmann  constants equal to $k_B=\hbar=1$. Later in the paper we shall use:
\begin{equation}
C={\pi T c \over 3v}, \quad  \mbox{as} \quad T\rightarrow 0 .
\end{equation}

 We are more interested in specific entropy  $\tt s$. We can integrate the 
equation and fix the integration constant from the third law of 
thermodynamics (${\tt s}=0$ at $T=0$). So for specific entropy we have the same low temperature behavior:
\begin{equation}
{\tt s}={\pi T c \over 3v}, \quad  \mbox{as} \quad T\rightarrow 0 .
\end{equation}
For quantum spin chains this formula agrees with \cite{tak} .
To formulate the problem precisely let us consider  Bose gas with delta interaction. The Hamiltonian of the model is:
\begin{equation}
H=\int_{-\infty}^{\infty} dx \left[\partial \psi_x^\dagger \partial \psi_x +g \psi^\dagger\psi^\dagger \psi \psi \right] .
\end{equation}

Here $\psi$ is a canonical Bose field and $g>0$ is a coupling constant.
The model was solved  in \cite{ll}. Physics of the model is described  in a  book \cite{bki}. 
First let us consider the model at zero temperature and in the infinite volume.
 The ground state is unique $|gs\rangle$. We consider gas with positive 
density.
We are interested in the entropy $S(x)$ of the part of the gas present on a
 space  interval $(0,x)$. Formally we can define it by means of the density
 matrix 
\begin{equation}
\rho =\mbox{tr}_\infty \left(|gs\rangle \langle gs| \right) \label{mdens}
\end{equation}
Here we traced out the 'external' degrees of freedom, they describe the gas
 on the rest of the ground state: on the unification of the intervals
  $(-\infty,0)$ and $(x, \infty )$. The density matrix $\rho$ describes
gas on the interval $(0,x) $.
Now we can calculate von Neumann entropy $S(x)$  of the part of the gas on the
 interval $(0,x)$ :
\begin{equation}
S(x)=-\mbox{tr}_{x} \rho \ln \rho \label{vn}
\end{equation}
Here we are taking the trace with respect to the degrees of freedom representing 
the part of the gas on the interval  $(0,x)$. In the major text books it is 
shown that the laws of thermodynamics can be derived from statistical
 mechanics, see for example \cite{lan,hu} .
Second law of thermodynamics states that the entropy is extensive parameter:
 the entropy of a subsystem $S(x)$
 is proportional to the system size  $x$:
\begin{equation}
S(x)={\tt s}x \quad \mbox{at} \quad T>0 \label{p1}
\end{equation}
Thermodynamics is applicable to the  subsystem of macroscopical size, 
meaning large $x$.
Here specific entropy $\tt s$ depends on the temperature.
For small temperatures the dependence simplifies, see (2):
\begin{equation}
S(x)={\pi Tc\over 3v }x, \quad x>{1\over T} \label{p2}.
\end{equation}

Let us try to find out how $S(x)$ depends on $x$ for zero temperature.
It is some functions of the size:
\begin{equation}
S(x)=f(x), \quad \mbox{at} \quad T=0 .
\end{equation}
Now let us apply the ideas of conformal field theory, see \cite{bpz,a,bcn} and also Chapter XVIII of \cite{bki} .
We can arrive to small temperatures from zero temperature by conformal mapping
$\exp{[2\pi T z/v]}$. It maps the whole complex plane of $z$ without the origin
to a strip of the width $1/T$. This replaces zero temperature by positive
 temperature $T$. 
The conformal mapping results in a replacement of  variable $x$ by 
$[v/\pi T]\sinh [\pi Tx/v  ]$.
So the entropy of the subsystem at temperature $T$ is given by the formula:
\begin{equation}
S(x)=f\left({v\over \pi T}\sinh \left [{\pi Tx\over v} \right]\right), \quad \mbox{at} \quad T>0 \label{co}
\end{equation}
In order to match this to formula (7) we have to consider asymptotic
of large $x$. The formula simplifies:
\begin{equation}
S(x)=f\left(\exp \left[ {\pi T(x-x_0)\over v} \right] \right),\quad Tx\rightarrow \infty.
\end{equation}

Here ${\pi Tx_0 / v}=-\ln  ({v / 2\pi T) }$ . 

Formulae (7) and (10) should coincide. Both represent the entropy of the subsystem for small positive temperatures.
This provides an equation for $f$:
\begin{equation}
f\left(\exp \left[ {\pi T(x-x_0)\over v} \right] \right)={\pi T c \over 3v }(x-x_0)
\end{equation}
This formula describes asymptotic for large $x$, so we  added $-x_0$ to the
 right hand side. We are considering the region $x>1/T$ and $x_0\sim \ln (1/T)$, so $x>>x_0$ at $T\rightarrow 0$.
In order  to solve the equation for $f$, let us introduce a new variable
$y=\exp \left[ {\pi T(x-x_0)/ v}\right] $ . Then the equation (11) reads:
\begin{equation}
f\left(y \right)={ c \over 3 }\ln y
\end{equation}
So we found the function $f$ in (8).
Now we know that  at zero temperature entropy of the gas containing on the 
interval $(0,x)$ is:
\begin{equation}
S\left(x \right)={ c \over 3 }\ln x , \quad \mbox{as} \quad x \rightarrow \infty \label{log}
\end{equation}
Let us remind that for Bose gas the central charge  $c=1$,  see \cite{bki}. Our result agrees with the third law
 of thermodynamics.
Specific entropy is a limit of the ratio $S(x)/x$ as $x\rightarrow \infty$.
The limit is  zero.
 
Now we can go back to our formula (\ref{co}) and substitute the expression for $f$, which we found:
\begin{equation}
 S(x)={c\over 3} \ln \left({v\over \pi T}\sinh \left [{\pi Tx\over v} \right]\right) \label{main}
\end{equation}
This formula describes crossover between zero and large temperature.

{ The proof presented here is universal.}  It is also applicable to quantum
 spin chains. For example to  XXZ spin chain.
The Hamiltonian of the model is:
\begin{equation}
{\bf H}=-\sum_{j}\big\{ \sigma^{x}_{j}
\sigma^{x}_{j+1}+\sigma^{y}_{j}\sigma^{y}_{j+1}
 + \Delta \sigma^{z}_{j}\sigma^{z}_{j+1}+2h\sigma^{z}_{j} \big\} \label{heis}
\end{equation}
It describes interaction of spins $1/2$. Here $\sigma$ are Pauli matrices.
The model can be described by a conformal field theory with  the central charge  $c=1$ in a  disordered regime: $-1\leq\Delta<1 $ and magnetic field $h$ is 
smaller then critical $h_c$,   see \cite{bki}. For magnetic field larger then 
critical the ground state  is a product of  wave functions of individual 
spins. Entropy of any  subsystem  is zero.

The  formula (13) was  discovered in \cite{hlw,vidal}. 
 More general formula can be found in \cite{keat, mez}.
The entropy of the subsystem $S(x)$
describes the entanglement of the subsystem and the rest of the ground state.

The formula (\ref{log}) describes the entropy of large subsystem
 $x\rightarrow \infty$ at zero temperature, first correction to
this formula is a constant term:

$$S\left(x \right)={ c \over 3 }\ln {x \over x_0}, \quad \mbox{as} \quad x \rightarrow \infty $$
The constant term defines the scale $x_0$. It does not depend on $x$, but it 
does depend on magnetic field and  anisotropy, see  \cite{jk}.

So far we discusses the entropy of entanglement for critical models. For
 non-critical [gap-full] models  asymptotic of $S(x)$ is a constant,
 depending on parameters of the model. For AKLT spin chain  $S(\infty)=2$,
see  \cite{fan}. For $XY$ spin chain $S(\infty)$ depends on anisotropy and 
magnetic field and shows singularities at phase transitions see \cite{its,pes}.

We evaluated von Neumann entropy (\ref{vn}). Calculation of generalized
 entropies [R\'enyi and Tsallis] can be reduced to calculation of a trace of
some power $\alpha$ of the density matrix:
$ tr_x(\rho^{\alpha})$
it was done in the paper \cite{jk}.

\section{Spin Chains with Arbitrary  Spin}
In this section we  consider higher values of spins.
We  calculate dependence of entropy of a subsystem on the value of interacting spin  $\bf s$. Let us consider isotropic XXX anti-ferromagnet \cite{heiz}. For spin
${\bf s}=1/2$ we can represent the Hamiltonian as:
\begin{eqnarray}
&{\bf H}_{\frac{1}{2}}=\sum_{n}X_n , \label{half}  \\
& X_n= \vec{S}_n \vec{S}_{n+1}=  S^{x}_{n}S^{x}_{n+1}+S^{y}_{n}S^{y}_{n+1}
 +  S^{z}_{n}S^{z}_{n+1} \label{scal}
\end{eqnarray}
The model is solvable by Bethe Ansatz \cite{bete}. 
A generalization of this model to spin $ {\bf s}=1$ was found in \cite{zamo}: 
\begin{equation}
{\bf H}_{{1}}=\sum_{n}\big\{ X_n - X^2_n \big\} \label{one}
\end{equation}
It was solved by  Takhatajan and   Babujian, see \cite{taht,bab}.
Generalization for higher spin $\bf s$ was found in \cite{fad,krs}
\begin{equation}
{\bf H}_{{s}}=\sum_{n} F( X_n)   \label{high}
\end{equation}
The function $F(X)$ is a polynomial of a degree $2{\bf s}$. It can be written  as 
\begin{equation}
F(X)=2\sum_{l=0}^{2{\bf s}} \sum_{k=l+1}^{2{\bf s}} \frac{1}{k}\prod_{\stackrel{\mbox{\small{$j=0$}}}{\mbox{\small{$j \neq l$}}}}^{2{\bf s}}\frac{X-y_j}{y_l-y_j} \label{dens}
\end{equation}
Here $y_l=l(l+1)/2-{\bf s}({\bf s}+1)$. The model (\ref{high}) also solvable by Bethe Ansatz.
There is no gap in the spectrum of this Hamiltonians.
The models can be described by a conformal field theory with  the central 
charge:  $$c=\frac{3{\bf s}}{{\bf s}+1},$$ 
see \cite{yaf}.
So the entropy of a  block of $x$ spins is:
\begin{equation}
S\left(x \right)={\frac{\bf s}{{\bf s}+1} }\ln {x }, \quad \mbox{as} \quad x \rightarrow \infty \label{spin}
\end{equation}
The coefficient in-front of the $\ln x$
increases from $1/3$ to $1$ as spin $\bf s$ increases from $1/2$ to $\infty$.

{\it Note}. It is interesting to remark that in ferromagnetic case
[different sign in from of the Hamiltonian] the entropy also scales logarithmically, but the
coefficient in front of the $\ln x$ is different, see \cite{sal,pop}.

\section{ Hubbard Model}

The Hamiltonian for the Hubbard model $H $ can be represented as
\begin{eqnarray}
& H & =-\sum_{j} \sum_{\sigma = \uparrow, \downarrow}  (c_{j,
\sigma}^\dagger c_{j+1, \sigma} + c_{j+1, \sigma}^\dagger c_{j,
\sigma})  \nonumber \\
+& u & \sum_{j} n_{j,\uparrow} n_{j,\downarrow} -
h\sum_{j}( n_{j,\uparrow}- n_{j,\downarrow})
\end{eqnarray} 

Here $c^\dagger_{j,\sigma}$ is a canonical Fermi operator on the lattice 
[operator of creation of an electron] and    $n_{j,\sigma}=c^\dagger_{j,\sigma}c_{j,\sigma}$ is an operator on number 
of electrons in cite  $j$ with spin $\sigma$.
Cite summation in the Hamiltonian goes through the whole infinite
 lattice. Coupling constant $u>0$ 
and magnetic field $h$ is below critical.
At half filled band only spin degree of freedom is gap-less, so at zero 
temperature $S\left(x\right)=(1/3)\ln x$.
Let us consider the model below half filling [less then one 
electron per lattice cite]. The model was solved in \cite{lw}.
Detailed description of physics of the model can be found in the text-book \cite{hub}.
Charge and spin separate in the model.
The model is gap-less. Both charge and spin degrees of freedom can be described
by Virasoro algebra with central charges
$c_c=1, \quad    c_s=1 .$
see \cite{fk}.
Also Fermi velocities are different for spin $v_s$ and charge $v_c$ degrees
 of freedom.
Both  spin and charge degrees of freedom contribute to the  specific entropy $s$.
So for small temperature we have:
${\tt s}=  {\pi T / 3v_s }+ {\pi T / 3v_c }.$
The entropy of the subsystem is proportional to the size of the system (second law):
\begin{equation}
S(x)=\left( {\pi T \over 3v_s }+ {\pi T \over 3v_c }\right)x
\end{equation}
Now we have to apply conformal arguments separately to spin and charge degrees of freedom. The results  add up:
\begin{equation}
S\left(x \right)={ 2 \over 3 }\ln x \label{eh}
\end{equation}
This describes the entropy of electrons on the interval $(0,x)$ in the infinite
 ground state at zero temperature.
 This is actually an asymptotic for large
 $x$. In case of the Hubbard model crossover formula [which mediate between
zero and positive temperature]  looks like this:
\begin{eqnarray}
 &S(x)= \label{ch} \\
&={1\over 3} \ln \left({v_s\over \pi T}\sinh \left [{\pi Tx\over v_s} 
\right]\right) + {1\over 3} \ln \left({v_c\over \pi T}\sinh \left [{\pi Tx\over v_c}\right]\right) \nonumber
\end{eqnarray}
The approach developed here is universal, see \footnote{Attractive case is considered in the  text-book \cite{hub}, page 630.} . It is applicable to other models of
 strongly correlated electrons, see \cite{ek}. For example formulae
 (\ref{eh}),(\ref{ch}) describe entropy in t-J  model below half-filling as
 well.

\section{Summary}

In the paper we described  universal properties of the entropy
in one dimensional gap-less models. We studied the entropy 
 of a  subsystem.  We considered scaling  of the entropy  in
 spin chains,  strongly correlated electrons
and  interacting Bose gas.
We  discovered crossover  formula for entropy of the subsystem, see
(\ref{ch}).
It describes entropy of a large subsystem at low temperature.
 We discovered a dependence of  entanglement 
on  spin $\bf s$ for antiferromagnet, see (\ref{spin}).


\section{Appendix}
Important characteristic of conformal field theory is a central charge.
It can be defined by considering an  energy-momentum tensor
 $T_{\mu, \nu}(z)$. Here $z$ is complex space-time variable $z=x+ivt$ and $v$ is a  Fermi velocity. The component of  energy-momentum tensor with $\mu =\nu =z$
is denoted by $T_{z,z}=T$. Correlation function of this operator has a singularity:
$\langle T(z) T(0)\rangle = {c/ z^4}$,
the coefficient $c$ is the  central charge.

\section{Acknowledgments}

We are grateful to A.Abanov, F.Essler, H.Frahm, P.A.Grassi, B.Q.Jin, 
A.Kluemper, S.Lukyanov and  B.McCoy for discussions. The paper was
 supported by NSF Grant PHY-9988566.

\end{document}